\documentclass[reprint,aps]{revtex4-1}

\usepackage{graphicx}
\usepackage{bm}
\newlength{\figwidth}
\usepackage{graphicx}
\usepackage{tabularx}
\usepackage{color}
\usepackage{amsmath}
\usepackage[rgb]{xcolor}
\usepackage{xcolor}

\begin{document}
\title{Phase-sensitive imaging of ferromagnetic resonance using ultrafast heat pulses}

\author{Feng Guo}
\affiliation{Cornell University, Ithaca, NY 14853, USA}

\author{J. M. Bartell}
\affiliation{Cornell University, Ithaca, NY 14853, USA}

\author{ D. H. Ngai}
\affiliation{Cornell University, Ithaca, NY 14853, USA}

\author{G. D. Fuchs}
\email{gdf9@cornell.edu}
\affiliation{Cornell University, Ithaca, NY 14853, USA}

\begin{abstract}

Measuring local magnetization dynamics and its spatial variation is essential for advancements in spintronics and relevant applications. Here we demonstrate a phase-sensitive imaging technique for studying patterned magnetic structures based on picosecond laser heating. With the time-resolved anomalous Nernst effect (TRANE) and extensions, we simultaneously image the dynamic magnetization and RF driving current density. The stroboscopic detection implemented in TRANE microscopy provides access to both amplitude and phase information of ferromagnetic resonance (FMR) and RF current. Using this approach, we measure the spatial variation of the Oersted driving field angle across a uniform channel. In a spatially nonuniform sample with a cross shape, a strong spatial variation for the RF current as well as FMR precession is observed. We find that both the amplitude and the phase of local FMR precession are closely related to those of the RF current.

\end{abstract}

\maketitle


  
	Improving the detection of local ferromagnetic resonance (FMR)  expands our ability to study magnetization dynamics and the underlying physics. From the application standpoint, appropriate measurement techniques are pivotal to develop and advance the next generation magnetic storage and memory technology. Here we present a study on local FMR measurement in conjunction with excitation current. We apply stroboscopic measurement techniques based on ultrafast heat pulses to detect both the RF current and FMR signal simultaneously. By measuring both absolute phase and amplitude, we establish the relation between the driving current and corresponding magnetic response.

   
	Several compelling techniques have been developed to study local magnetization dynamics, including micro-focused Brillouin light scattering\cite{DemokritovD_BLS_Transmag08, NembachSSJKMK_BLS_prb11, stamps2012solid, Urazhdindukklwd_BLS_nnano14}, force-based FMR detection\cite{KleinDNBGHLSTV_fmrfm_prb08, LeeOXHYBPH_fmrfm_nature10, ChiaGBM_localization_prl12, ChiaGBM_array_apl12, GuoBM_edge_prl13, AdurDWMBZPYH_prl14, HamadehDHMBMNVACDPMDK_fmrfmYIG_prl14}, time-resolved Kerr microscopy\cite{HiebertSF_kerr_prl97, ParkEEBC_kerr_prl02, HillebrandsO_kerr_book03, KeatleyGDHCK_kerr_apl11}, and X-ray magnetic circular dichroism\cite{ChembroluSYTTKCSA_xray_prb09, MarchamKNHCSVTCKSA_xray_jap11, VogelKMDCTSVM_xray_prl11}, to name a few. Very recently, time-resolved anomalous Nernst effect (TRANE) microscopy has been developed for magnetic imaging as well as for stroboscopic FMR measurement\cite{BartellDLF_TRANE15}. 
	
	Relevant to this work, spin torque ferromagnetic resonance (ST-FMR)\cite{TulapurkarSFKMTDWY_ST-FMR_nature05, SankeyBGKBR_ST-FMR_prl06, LiuTRB_ST-FMR_prl11} is a phase-sensitive technique that has been effective for studying spin Hall effect physics. The rectified DC signal measured with ST-FMR is sensitive to the relative phase between the magnetization precession and the RF current, while TRANE microscopy probes the absolute precession phase. Also, ST-FMR lacks the ability to probe the spatial variations that might occur in the devices. In addition to the existing electrical measurements, a phase-sensitive FMR measurement technique using Magneto-optic Kerr effect has also been recently reported\cite{MoriyamaSM_fmrKerr_jap15}.

  
	Here we introduce a method of phase-sensitive magnetic imaging based on TRANE microscopy, combining spatial scanning and phase detection capabilities.  We demonstrate simultaneous detection of local spin wave resonance and RF current. This capability enables imaging of the magnetic dynamic susceptibility in the GHz range. A distinct feature of this work is that we demonstrate a technique for measuring the local amplitude and phase of both magnetic precession and microwave excitation current. This enables us to image the spatial variations of the magnetic dynamics that are lost in other electrical measurement techniques. The relationship between excitation and response is relevant in understanding the origin of the torques that drive magnetic dynamics.


We first describe the essential measurement procedure and then explain the detection method of both the RF driving current and the FMR response. Next we quantitatively analyze the FMR phase in response to a varied RF current phase and show that the phase-sensitive FMR spectra measured in a uniform current channel reveal the local driving field orientation. Using phase dependent imaging, we also demonstrate that a spatially nonuniform channel shows a strong spatial variation, both for FMR and for RF current.


\begin{figure*}[tb]
  \centering
  \graphicspath{{./}{Figures/}}
  \includegraphics[width=\textwidth]{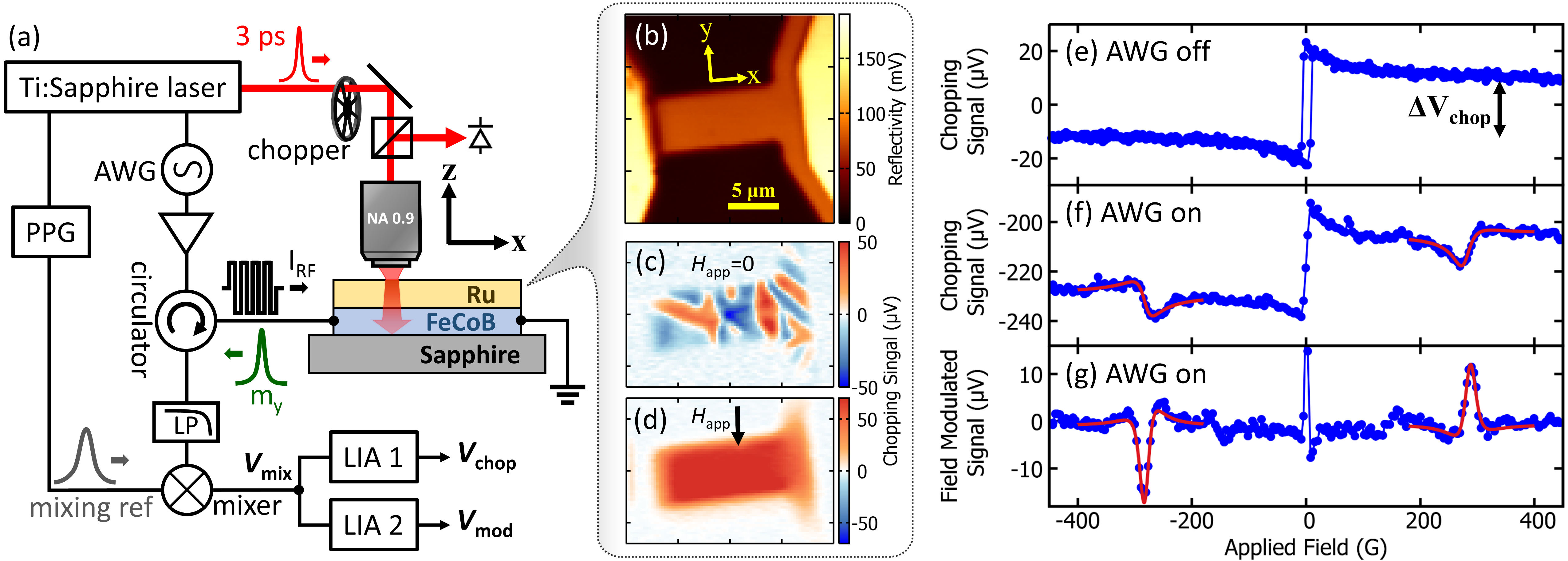}
  \caption{(Color online) (a) Schematics of the time-resolved anomalous Nernst effect (TRANE) setup. (b) Reflected laser intensity shows a micrograph of the sample. Without any applied current, the chopping referenced signal measures the y-component of the magnetization, $m_y$, in a demagnetized state at zero field (c) showing domain patterns and a saturated state (d) under a large applied field. (e) Hysteresis loop measured with the chopping signal with an in-plane applied field (5$^\circ$ away from the sample length direction). No current nor modulation field are applied in this measurement. (f) A similar hysteresis loop measurement with a 5.7~GHz RF driving current. Ferromagnetic resonance signal is seen for both applied field directions, near $\pm 280$~G. The large constant background voltage is due to the combination of local heating and the RF current. (g) The field modulated signal, only sensitive to the magnetic response, is measured simultaneously with the chopping signal in (f). The red curves in (f) and (g) are the fits for ferromagnetic resonance. All the data shown in (e-g) are the mixed signals locked into the chopping reference ($V_{\rm chop}$) and the field modulating reference ($V_{\rm mod}$).
    \label{fig:setup}}
\end{figure*}

As its name suggests, the heart of TRANE microscopy is the anomalous Nernst effect (ANE)\cite{SlachterBV_ANE_prb11, WeilerACHWOIRTGGS_ANE_prl12, vonBierenBGA_ANE_apl13, WuHPB_ANE_apl14, LeeKYKLLJLSSSP_ANE_15}: an electric field, $E_{\rm ANE}=-N \mu_0 {\bf m} \times \nabla T$, produces an ANE voltage $V_{\rm ANE}$ associated with the magnetization ${\bf m}$, through the anomalous Nernst coefficient $N$ and the temperature gradient $\nabla T$. We use a hybrid measurement scheme that combines optical generation of a pulsed thermal gradient and electrical detection of an ANE voltage, in order to stroboscopically detect the transient local magnetization. We point out that the spatial and temporal resolution of the TRANE microscopy are ultimately determined by the spatial and temporal profiles of the thermal gradient. With a $\nabla T$ along the z direction and a pair of contacts along the x direction (Fig.~\ref{fig:setup}), the measured ANE voltage is sensitive to the y component of the local magnetization, $m_y$.

Fig.~\ref{fig:setup}(a) depicts the schematics of the TRANE setup. A vertical thermal gradient is generated by a 780~nm Ti:Sapphire laser with 3~ps long pulses and a 25.3~MHz repetition rate. The laser intensity is also modulated at 100~kHz using a polarizer and a  photoelastic modulator. To create an RF driving field we use an arbitrary waveform generator (AWG) that applies a continuous waveform RF current to the sample via a circulator. The laser and AWG are synchronized such that there is a constant phase relation between the RF current and the laser pulse train, which allows us to stroboscopically probe the instantaneous magnetization of the spin waves. Each laser pulse generates a  pulsed signal, and the voltage pulse is demodulated in a mixer by combining it with a 1.5~ns duration electrical reference pulse that enters the mixer at the same time. The mixed output voltage is then measured by lock-in amplifiers.


In the following, we first discuss the various origins of the signal, followed by the measured spectra that contain both magnetic and RF current information. There are two signals generated by the laser pulses. Besides the above-mentioned magnetic term from the ANE voltage, an increase in the sample resistance $\Delta R_{\rm heat}$ induced by local laser heating also contributes to the total voltage pulse generated across the sample:
\begin{equation}
  \label{eq:Vchopping}
  V_{\rm sample}=V_{\rm ANE}+V_{\rm J}.
\end{equation}
Here the second term, $V_{\rm J}=-I(t)\Delta R_{\rm heat}(t)$\footnote{
The minus sign is merely due to the choice of the ground. We define the sign of the current to be consistent with that of the driving field $h_{\rm RF}$, i.e., when $I_{\rm RF}>0$, $h_{\rm RF}>0$.}, 
is determined by the instantaneous local current following through the heated volume: $I(t)=I_{\rm RF}^0\sin(\omega t +\varphi_{\textsc{\tiny RF}})$, in which $I_{\rm RF}^0$ is the local RF current amplitude, $\omega$ is the current frequency, and $\varphi_{\textsc{\tiny RF}}$ is the RF current phase. As will be described later, we use $V_{\rm J}$ to measure the phase and spatial profile of the RF current. After the mixer, the voltage pulse $V_{\rm sample}$ from the sample is converted to a mixed signal $V_{\rm mix}$. A lock-in amplifier is used to measure the signal with respect to the chopping reference, which we refer to as $V_{\rm chop}$. Furthermore, to reject the non-magnetic background we also apply a 350~Hz, 7~G modulation field for measuring FMR signal. The field modulated signal $V_{\rm mod}(H)$, which is proportional to $\partial V_{\rm mix}(H)/\partial H$, is measured as a function of the applied field while recording the FMR spectra.

The samples consist of ${\rm Fe_{60}Co_{20}B_{20} (4~nm)/Ru (4~nm)}$ bilayers, deposited on the sapphire substrate as a heat sink. The bar samples have a dimension of ${\rm 5~\mu m \times 12~\mu m}$ and have a resistance of about $300~\Omega$. We chose this simple bilayer structure to minimize the potential spin Hall effect, as confirmed by a separate ST-FMR experiment ($J_s/J_c = 0.015\pm 0.009$ which is several times smaller than the reported values for platinum
\cite{AndoTHSIMS_PtSpinHall_prl08, LiuBR_PtSpinHall_arxiv11, AzevedoVRLR_PtSpinHall_prb11, LiuTRB_ST-FMR_prl11, ZhangHKYP_PtSpinHall_nature15}). The Oersted field has a known spatial profile determined by the current. Therefore using the Oersted field as the only driving torque simplifies the data interpretation and helps us to establish the phase analysis.

Examples of measured spectra are shown in Figs.~\ref{fig:setup}(e-g).  With the AWG off, the chopping signal contains only the ANE signal. Fig.~\ref{fig:setup}(e) shows a hysteresis loop, with an in-plane field aligned $5^\circ$ off from the length of the bar (x-direction). The voltage difference between magnetization saturated in opposite directions, $\Delta V_{\rm chop}$, corresponds to the y-component of the saturation magnetization: $2M_s \sin5^\circ$. However, when the AWG is turned on, a 5.7~GHz RF current creates a large background due to the contribution from $V_{\rm J}$. This constant background is determined by the fixed phase of the RF current with respect to the laser stroboscope. As we will discuss later, the voltage background in the chopping signal indeed depends on the AWG phase. Nevertheless, the signal due to magnetic reversal $\Delta V_{\rm chop}$ remains the same, as shown in Fig.~\ref{fig:setup}(f). As a result of the RF excitation current, an FMR precession signal is also observed for both field directions. By the comparing the FMR signal to the $\Delta V_{\rm chop}$, we calculate the precession angle to be $(1.5\pm0.1)^\circ$
\footnote{To obtain the precession angle $\theta_{\rm p}$, we use $\sin \theta_{\rm p}/\sin 5^\circ=2 V_{\rm FMR}/\Delta V_{\rm chop}$, where $5^\circ$ is the titling angle of the in-plane applied field with respect to x direction, $V_{\rm FMR}$ is the voltage amplitude of the FMR signal, and $\Delta V_{\rm chop}$ is the voltage difference between opposite large fields in the hysteresis loop [Fig.~\ref{fig:setup}(f)].}.
 Finally, to isolate the magnetic signal from the non-magnetic RF current contribution, a field modulated signal $V_{\rm mod}(H)$ is recorded simultaneously shown in Fig.~\ref{fig:setup}(g). Only the FMR signal is revealed by locking into the field modulation, along with a peak near zero field due to the magnetization reversal.


\begin{figure}[tb]
  \centering
  \graphicspath{{./}{Figures/}}
  \includegraphics[width=\figwidth]{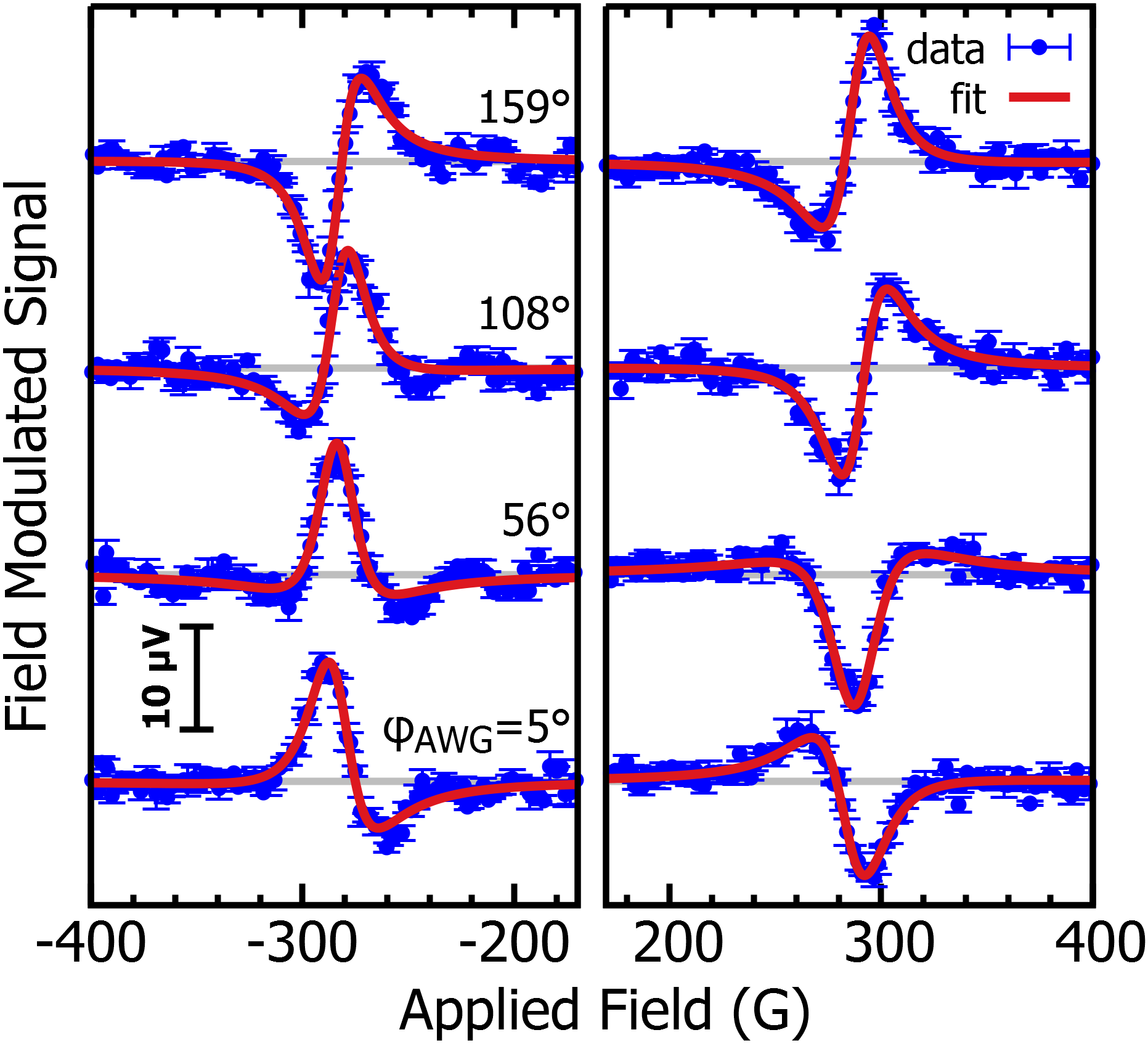}
  \caption{Examples of field modulated FMR spectra measured as a function of AWG phase, $\varphi_{\textsc{\tiny AWG}}$, for both negative (left) and positive (right) field directions.
    \label{fig:AWGphasespectra}}
\end{figure}

In this section, we focus on characterizing the precession phase of the measured FMR spectra. We measure the FMR precession phase, $\varphi_{\textsc{\tiny FMR}}$, through the line shape of the spectra. The field modulated spectrum is a linear combination of the real ($\chi'$) and imaginary ($\chi''$) dynamic susceptibilities, given by\footnote{
Note that Eq.~\ref{eq:FMRphase} is valid under small modulation field approximation. In our setup, we use a modulation field of about 7~G and the full width at half maximum of the linewidth is typically around 33~G. Thus the expression in Eq.~\ref{eq:FMRphase} is adequate for fitting the spectra.
}:
\begin{equation}
  \label{eq:FMRphase}
  V_{\rm mod}(H) \propto  \frac{d\chi'(H)}{dH} \cos \varphi_{\textsc{\tiny FMR}}+ \frac{d\chi''(H)}{dH} \sin \varphi_{\textsc{\tiny FMR}}.
\end{equation}
The precession phase $\varphi_{\textsc{\tiny FMR}}$ is directly measured from the FMR spectrum, and it depends on the phase of the RF current at the time of the stroboscopic probe, as we will discuss in the following.

To vary the RF current phase, we use the AWG to tune the relative phase of the output waveform which we define as $\varphi_{\textsc{\tiny AWG}}$. Nevertheless, $\varphi_{\textsc{\tiny RF}}\neq \varphi_{\textsc{\tiny AWG}}$ in general since there is an initial current phase randomly determined upon AWG triggering ($\varphi_{\textsc{\tiny AWG}}^0$). Once the AWG is triggered and synchronized with the laser pulses, $\varphi_{\textsc{\tiny AWG}}^0$ remains constant throughout the measurements, and it can be determined as shown later. Thus the resultant RF current phase is:
\begin{equation}
  \label{eq:RFcurrent}
   \varphi_{\textsc{\tiny RF}}=\varphi_{\textsc{\tiny AWG}}-\varphi_{\textsc{\tiny AWG}}^0.
\end{equation}

\begin{figure}[tb]
  \centering
  \graphicspath{{./}{Figures/}}
  \includegraphics[width=\figwidth]{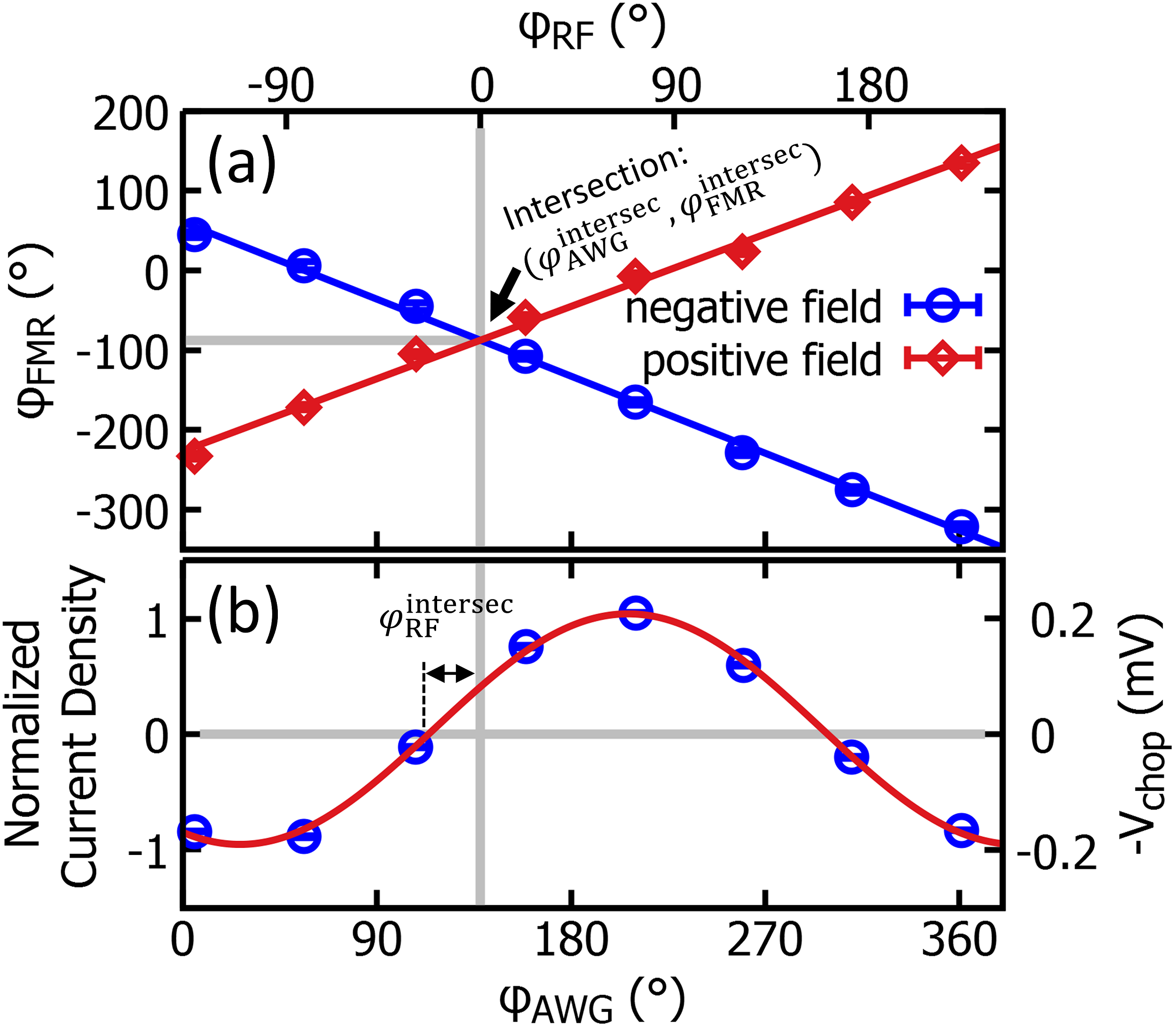}
  \caption{(a) FMR precession phases ($\varphi_{\textsc{\tiny FMR}}$) of both positive (diamonds) and negative (circles) field directions as functions of an increasing AWG phase ($\varphi_{\textsc{\tiny AWG}}$). $\varphi_{\textsc{\tiny FMR}}$ is measured from fitting the spectra such as those shown in Fig.~\ref{fig:AWGphasespectra}, with the laser placed at the center of the sample. The intersection of the positive and negative field curves is located at: $\varphi_{\textsc{\tiny AWG}}^{\rm intersec}=(137.9\pm4.0)^\circ$ and $\varphi_{\textsc{\tiny FMR}}^{\rm intersec}=(-87.4\pm4.1)^\circ$. (b) The normalized RF current density is measured as a function of $\varphi_{\textsc{\tiny AWG}}$ through the chopping reference voltage. The red curve is a sinusoidal fit. The intersection in (a) corresponds to an RF current phase of  $\varphi_{\textsc{\tiny RF}}^{\rm intersec}=(20.6\pm4.2)^\circ$, which is the difference in the current phase measured in (a) and (b).
    \label{fig:RFphaselag}}
\end{figure}

Fig.~\ref{fig:AWGphasespectra} shows the $\varphi_{\textsc{\tiny AWG}}$ dependent FMR spectra, measured at the center of the bar sample for both positive and negative applied fields. For quasi-uniform FMR mode, we adopt a macrospin model using Landau-Lifshitz-Gilbert equation with an oscillating Oersted driving field. The FMR precession phases at positive ($\varphi_{\textsc{\tiny FMR}}^+$) and negative ($\varphi_{\textsc{\tiny FMR}}^-$) fields can be written as:
\begin{subequations}
\label{eq:FMRphasevsAWGphase}
\begin{align}
        \varphi_{\textsc{\tiny FMR}}^+&=\varphi_{\textsc{\tiny RF}}-90^\circ+\theta_{\rm Oe},\\
        \varphi_{\textsc{\tiny FMR}}^-&=-\varphi_{\textsc{\tiny RF}}-90^\circ+\theta_{\rm Oe},
\end{align}
\end{subequations}
where $\theta_{\rm Oe}$ is the angle of the Oersted field with respect to the sample plane. The sign change under magnetic field reversal results from the precession orientation; the term $-90^\circ$ originates from the fact the magnetic response is $90^\circ$ behind the driving field (note that at resonance $\chi''=0$).

\begin{figure}[tb]
  \centering
  \graphicspath{{./}{Figures/}}
  \includegraphics[width=\figwidth]{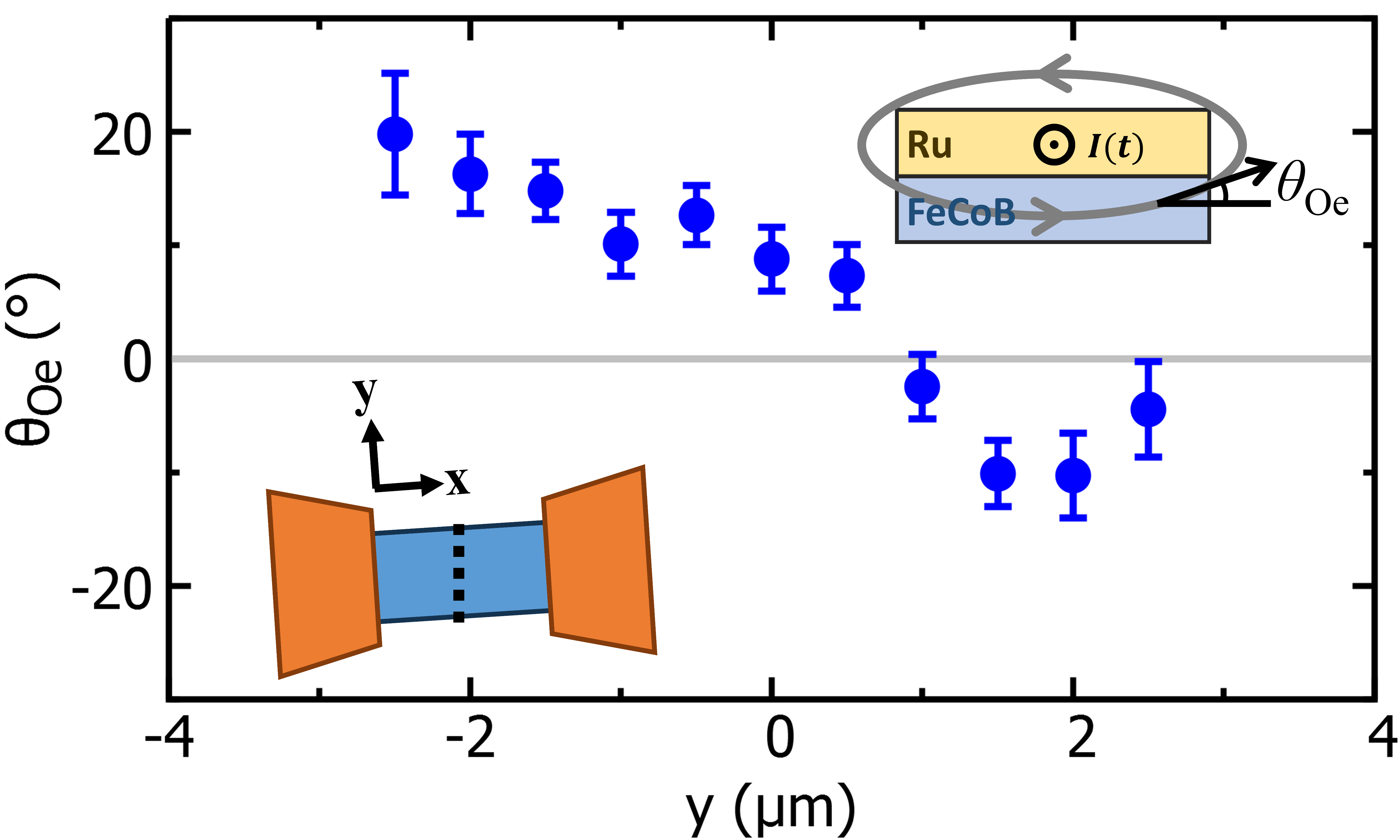}
  \caption{Measured Oersted field angle, $\theta_{\rm Oe}$, as a function of laser's y position. The upper inset is the schematics for the Oersted field angle, and the lower inset illustrates that the laser scans along the width of the channel (dotted line) for measuring $\theta_{\rm Oe}$.
    \label{fig:Oevsy}}
\end{figure}

At the center of the bar structure, we expect an in-plane Oersted driving field ($\theta_{\rm Oe}=0$). After including the initial AWG phase, the intersection of Eqs.~\ref{eq:FMRphasevsAWGphase} (a) and (b), ($\varphi_{\textsc{\tiny AWG}}^{\rm intersec}, \varphi_{\textsc{\tiny FMR}}^{\rm intersec}$), locates at: $\varphi_{\textsc{\tiny AWG}}^{\rm intersec}=\varphi_{\textsc{\tiny AWG}}^0$ and $\varphi_{\textsc{\tiny FMR}}^{\rm intersec}=-90^\circ$. The measured FMR phase from Fig.~\ref{fig:AWGphasespectra} as a function of AWG phase (also the calculated RF current phase using the measured $\varphi_{\textsc{\tiny AWG}}^0$
\footnote{The value of AWG phase at intersection suggests  $\varphi_{\textsc{\tiny AWG}}^0=\varphi_{\textsc{\tiny AWG}}^{\rm intersec}=(137.9\pm4.0)^\circ$.})
 is summarized in Fig.~\ref{fig:RFphaselag}(a). The fitted slope for positive (negative) field is $1.01\pm0.03$ ($-1.07\pm0.02$), consistent with the prediction of Eq.~\ref{eq:FMRphasevsAWGphase}. At the point of intersection, $\varphi_{\textsc{\tiny FMR}}^{\rm intersec}=(-87.4\pm4.1)^\circ$, in agreement with the predicted value of $-90^\circ$.

RF current phase is also separately measured with the chopping reference signal, $V_{\rm chop}$, shown in Fig.~\ref{fig:RFphaselag}(b). A sinusoidal waveform is seen as expected. As a subtle point, the intersection in Fig.~\ref{fig:RFphaselag}(a) does not exactly align with the zero phase in Fig.~\ref{fig:RFphaselag}(b), as illustrated by the gray lines. Instead, a discrepancy of $20.6^\circ$ in the RF current phase is found. In the following we explain the discrepancy by a difference in the temporal evolution of the $V_{\rm ANE}$ and $V_{\rm J}$ pulses that contribute to the signal. The phase measured with $V_{\rm ANE}$ voltage pulse depends on the temporal profile of the thermal gradient. In contrast, the $V_{\rm J}$ pulse due to heating is determined by the temperature change. Finite element simulation suggests that the absolute temperature has a slightly slower response to the laser pulse than the thermal gradient\cite{BartellDLF_TRANE15}. In addition, the temporal profile of the temperature has a slower delay. As a result, the measured RF current phase in Fig.~\ref{fig:RFphaselag}(b) has a forward phase shift compared to that measured with magnetic precession phase  [Fig.~\ref{fig:RFphaselag}(a)]. Finally, a separate fast mixing experiment (using a narrower $\sim$80~ps  reference pulse) also confirms a small but measurable delay ($\lesssim$20~ps) between the absolute temperature and thermal gradient pulses.

Now we use the FMR phase relation established earlier in Eq.~\ref{eq:FMRphasevsAWGphase} to measure the spatial variation of the Oersted field. To do so, we use the sum of Eq.~\ref{eq:FMRphasevsAWGphase}(a) and \ref{eq:FMRphasevsAWGphase}(b) to obtain the Oersted field orientation:
\begin{equation}
  \label{eq:Oe}
  \theta_{\rm Oe}=(\varphi_{\textsc{\tiny FMR}}^+ +\varphi_{\textsc{\tiny FMR}}^- +180^\circ)/2.
\end{equation}
Note that the expression for $\theta_{\rm Oe}$ is independent of $\varphi_{\textsc{\tiny RF}}$, and hence the measured $\theta_{\rm Oe}$ is not affected by the random initial AWG phase $\varphi_{\textsc{\tiny AWG}}^0$. The Oersted field direction is measured as a function of the y position of the laser, as presented in Fig.~\ref{fig:Oevsy}. Though scattered, the data show a general trend that is consistent with the expected Oersted field distribution in the sample. Near the center of the structure, the Oersted field direction is mostly in-plane: $\theta_{\rm Oe}\approx 0$; while approaching either edges, the Oersted field tilts out of the plane, towards either the positive or negative z directions.


\begin{figure}[tb]
  \centering
  \graphicspath{{./}{Figures/}}
  \includegraphics[width=\figwidth]{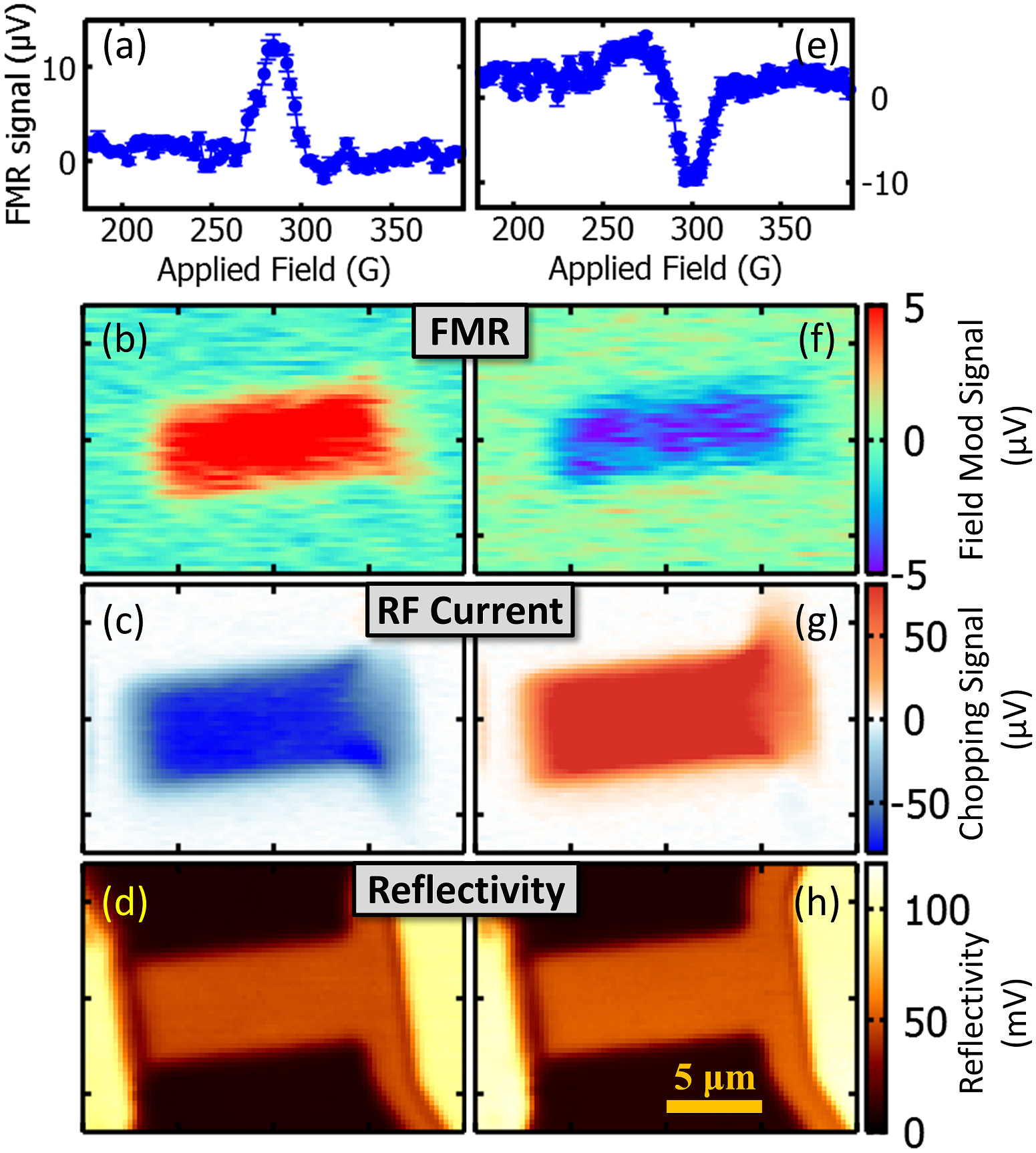}
  \caption{(Color online) Field modulated FMR spectra with $\varphi_{\textsc{\tiny AWG}}=260^\circ$ and $5^\circ$, respectively, showing either positive (a) or negative (e) signal at resonance. For the spectrum in (a), the applied field is fixed at 285~G where the peak locates while the FMR signal (b), RF current (c) and reflectivity (d) is measured simultaneously. (f-h) Similar imaging is recorded for the spectrum in (e) while the applied field is fixed at 300~G.
    \label{fig:barimaging}}
\end{figure}

Next, we demonstrate the scanning capability with phase-sensitive imaging of both FMR signal and RF current. We select two values of $\varphi_{\textsc{\tiny AWG}}$ from the data in Figs.~\ref{fig:AWGphasespectra} and \ref{fig:RFphaselag}, $260^\circ$ and $5^\circ$, that respectively have a positive and negative resonance peak in the field modulated signal. Fig.~\ref{fig:barimaging} shows the imaging of FMR signal ($V_{\rm mod}$), RF current signal ($V_{\rm chop}$), and reflectivity for the each AWG phase. Both FMR signal and RF current signal change sign between the two phases, which is consistent with the previous results in Fig.~\ref{fig:RFphaselag}. (Figs.~\ref{fig:RFphaselag} and \ref{fig:barimaging} have the same $\varphi_{\textsc{\tiny AWG}}^0$.) Regardless of its phase, we find that the RF current flows uniformly within the micrometer scale bar structure, unlike the case of millimeter scale channels where the RF current could be spatially varying\cite{VlaminckSPFBH_spatialvariation_apl12}; while the quasi-uniform FMR signal appears to have a relatively broad distribution with a smooth variation near the edges.


\begin{figure}[tb]
  \centering
  \graphicspath{{./}{Figures/}}
  \includegraphics[width=\figwidth]{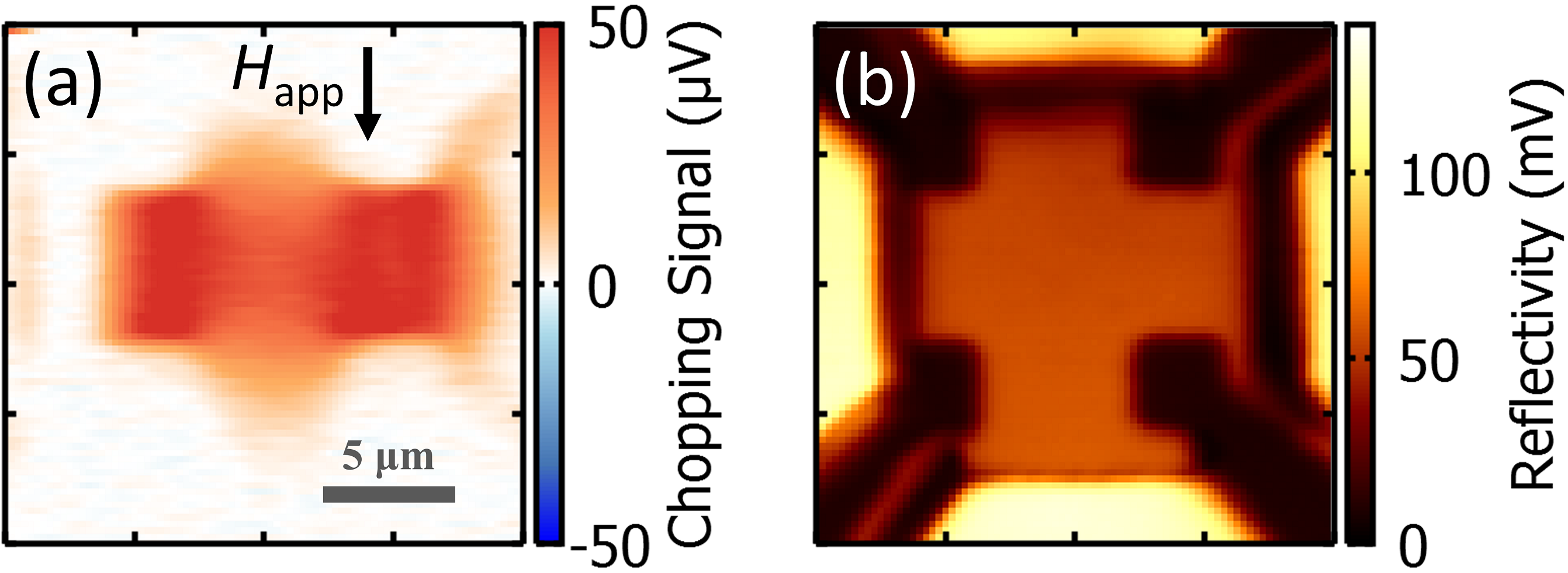}
  \caption{(Color online) (a) ANE imaging of the cross sample in the saturated state. A large magnetic field is applied along the y direction, with zero current applied. A pair of electrical contacts are connected to the left and right pads of the cross. (b) The reflectivity of the laser is measured along with the magnetic imaging.
    \label{fig:cross_saturated}}
\end{figure}

So far we have discussed the sample with a straight channel, in which case the RF current is uniformly distributed and maintains constant phase inside the sample. In the following we perform TRANE measurements on a nonuniform channel with a cross geometry. Although the cross displays a slightly more complicated scenario where both the amplitude and phase of the RF driving current is nonuniform, it better demonstrates TRANE's imaging capability for both the current and magnetic response. Fig.~\ref{fig:cross_saturated}(a) shows the magnetic imaging of the cross sample saturated in y direction, without the applied current. When measuring the cross structure, we connect the left and right contact pads to the RF current source, with the top and bottom pads left open. The measured $V_{\rm chop}$ remains sensitive to $m_y$ under this configuration.  Instead of a uniform magnetic signal shown in Fig.~\ref{fig:setup}(d) for the bar sample, the cross sample has a weaker signal at the center than that near the left and right pads.  This is because when the focused laser spot is located in the middle of the cross, the local $V_{\rm ANE}$ is shunted through the top and bottom arms of the cross\cite{BartellDLF_TRANE15}, resulting in a lower voltage than what is measured in either the left or right arm. Therefore, although the $V_{\rm ANE}$ of the saturated magnetization is expected to be uniform, the nonuniform magnetic imaging in Fig.~\ref{fig:cross_saturated}(a) is merely a result of the spatial dependence of the detection efficiency due to the sample geometry.

\begin{figure}[tb]
  \centering
  \graphicspath{{./}{Figures/}}
  \includegraphics[width=\figwidth]{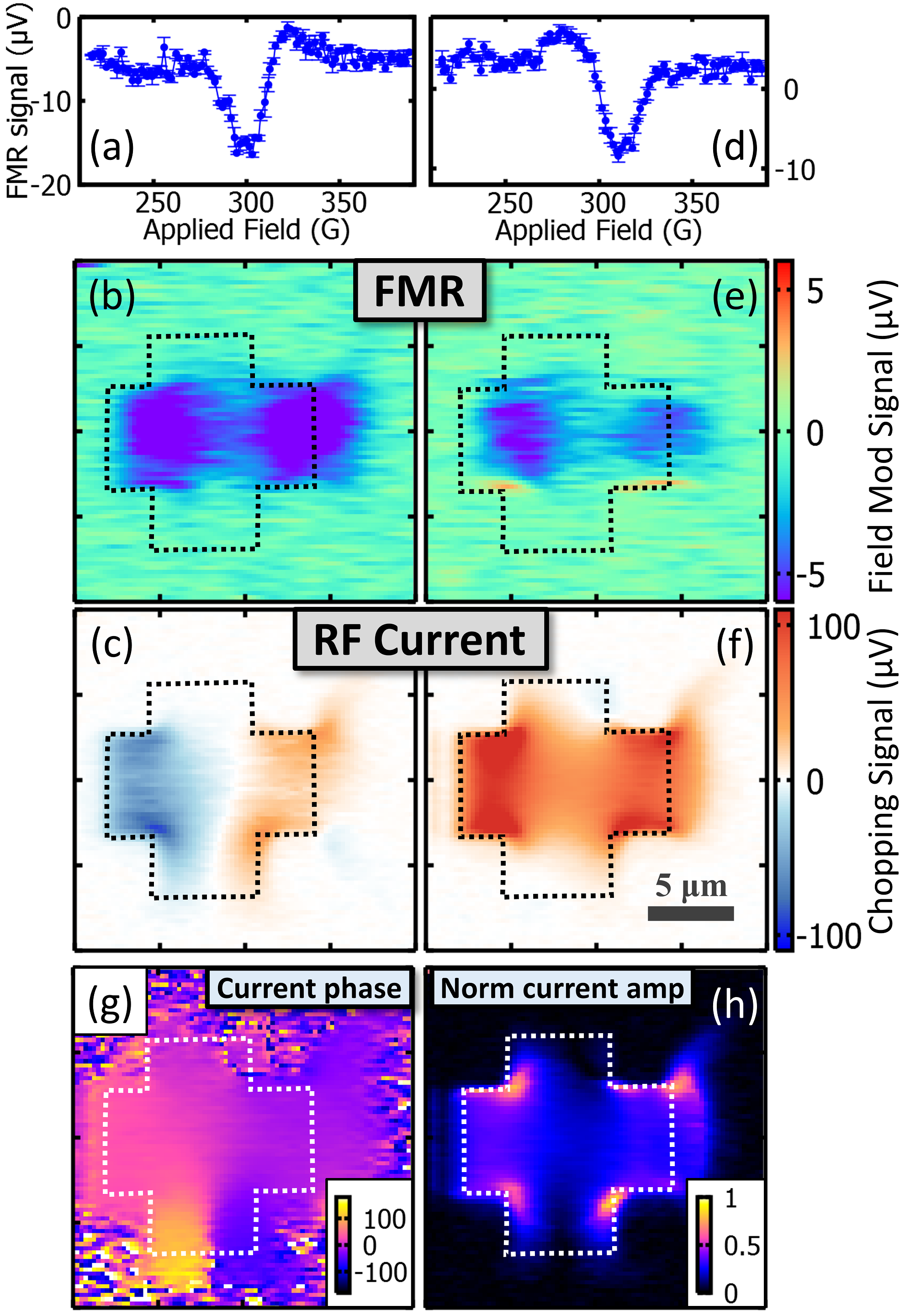}
  \caption{(Color online) (a) A FMR spectrum with $\varphi_{\textsc{\tiny AWG}}=310^\circ$ measured at the right arm of the cross. (b) and (c) show the FMR and RF current images, respectively, for the spectrum in (a) when the applied field is fixed at 300~G. (b) Another FMR spectrum $\varphi_{\textsc{\tiny AWG}}=210^\circ$, also measured at the right arm of the cross, and its corresponding FMR (e) and RF current (f) imaging with the applied field fixed at 310~G. The images of current phase (g) and normalized current intensity (h) are reconstructed from (c) and (f). The effect of the spatial dependence of the detection efficiency [Fig.~\ref{fig:cross_saturated}(a)] has been removed in the current intensity map. The dashed contours of the cross are obtained from the simultaneous reflectivity measurements.
    \label{fig:crossimaging}}
\end{figure}

We now apply the RF current to investigate FMR imaging for the cross structure. Two different RF current phases are used, and for each current phase both the FMR signal and the RF current are imaged. The results for $\varphi_{\textsc{\tiny AWG}}=310^\circ$ are shown in Figs.~\ref{fig:crossimaging}(a-c), and the similar measurements are done for $\varphi_{\textsc{\tiny AWG}}=210^\circ$ shown in Figs.~\ref{fig:crossimaging}(d-f). The most notable result is the imaging of Fig.~\ref{fig:crossimaging}(c) in which the current signal changes the sign across the sample. The RF current signal for $\varphi_{\textsc{\tiny AWG}}=210^\circ$ does not change sign although it does go through a phase shift. The FMR response shown in Figs.~\ref{fig:crossimaging}(b) and (e) also has a strong spatial variation, and it can even go through a sign change for particular AWG phases (not shown here). Lastly, by combining two current images [Figs.~\ref{fig:crossimaging}(c) and (f)] measured at different AWG phases, we can reconstruct the images for both the phase and the amplitude of the RF current, shown in Figs.~\ref{fig:crossimaging}(g) and (h) respectively. We point out that we use the normalized current distribution in Fig.~\ref{fig:crossimaging} (c) to remove the effect of the spatially nonuniform detection efficiency function indicated in Fig.~\ref{fig:cross_saturated}(a).

\begin{figure}[tb]
  \centering
  \graphicspath{{./}{Figures/}}
  \includegraphics[width=\figwidth]{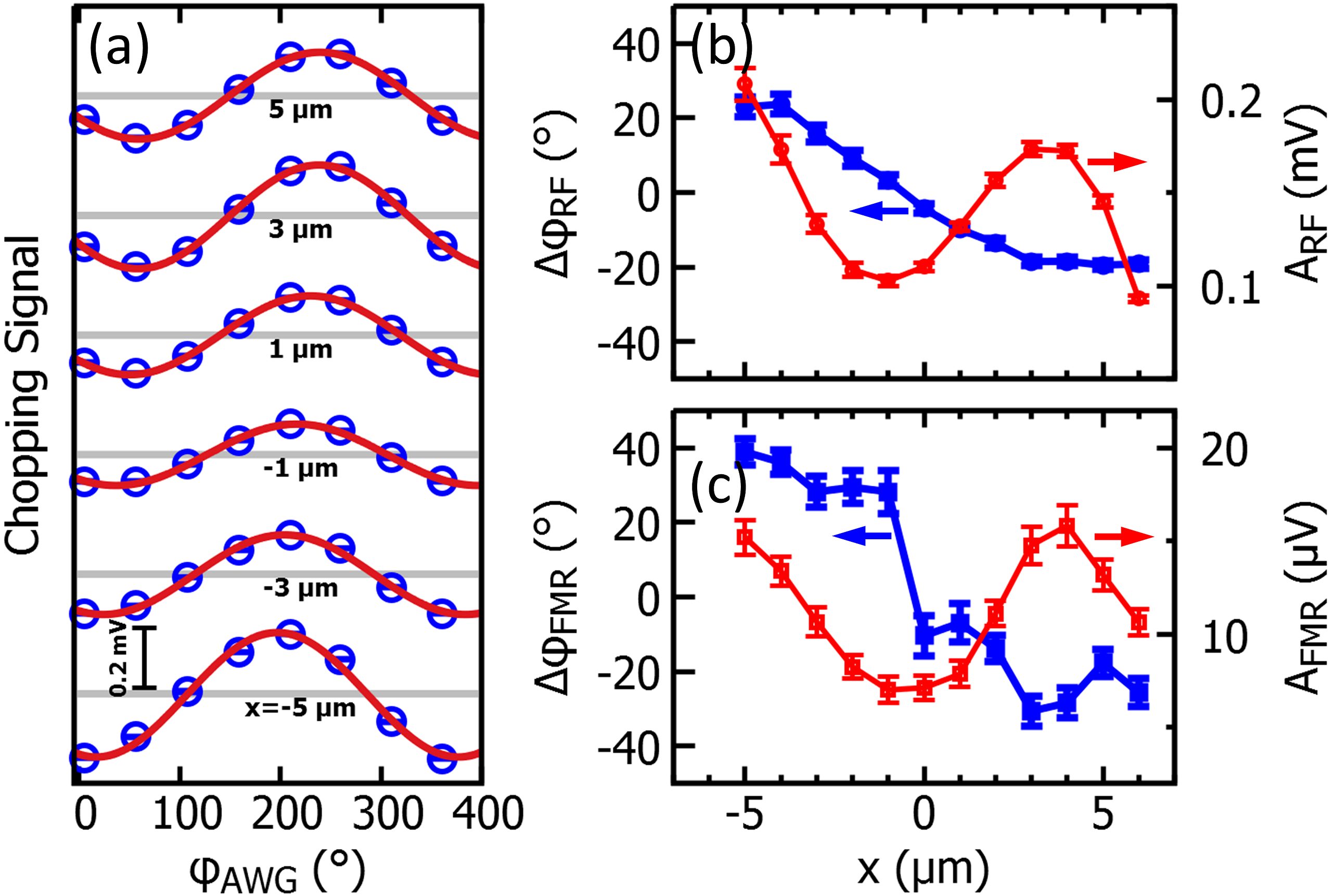}
  \caption{ (a)  RF current signal as a function of AWG phase, measured at various x positions. $x=0$ corresponds to the middle of the cross. The relative phase (solid blue) and amplitude (hollow red) for the RF current (b) and field modulated FMR signal (c) are also measured as functions of x.
    \label{fig:cross_phasevsx}}
\end{figure}

To further investigate the features imaged in the cross structure that are distinct from the bar structure, we measure the phase and the amplitude for both the RF current and FMR response at points along the x direction across the middle of the sample. The results are shown in Fig.~\ref{fig:cross_phasevsx}. Note that the $\varphi_{\textsc{\tiny AWG}}$ in Fig.~\ref{fig:cross_phasevsx} is consistent with that in Fig.~\ref{fig:crossimaging}, and the RF current sign change for $\varphi_{\textsc{\tiny AWG}}=310^\circ$ is also observed in Fig.~\ref{fig:cross_phasevsx}(a). As illustrated in Fig.~\ref{fig:cross_phasevsx}(a), not only the amplitude but also the phase of RF current varies with x position. The current amplitude reduces at the cross center, which can be understood from current spreading and signal shunting previously observed in Fig.~\ref{fig:cross_saturated}(a). However, the current phase varies monotonically across the sample, plotted in Fig.~\ref{fig:cross_phasevsx}(b). We attribute the phase shift of the current to the shape dependent inductance. As the current follows along the sample it encounters a geometry induced inductance variation, particularly at the center, which alters the current phase. In company with the driving current, the FMR phase also decreases along x, shown in Fig.~\ref{fig:cross_phasevsx}(c). The amplitude of the FMR signal is also closely related to the RF current amplitude. We conclude that for these samples, where the sample dimension is much longer than the magnetic exchange length, the spatially dependent phase and amplitude of FMR precesion is strongly influenced by the local excitation. 


In summary, we have demonstrated simultaneous measurements of local FMR and RF current using TRANE microscopy and its extensions. We have studied samples driven solely by Oersted fields to establish a quantitative phase relation between excitation current and magnetic response at GHz frequency, which is useful for future research of spin torque devices. We have also shown stroboscopic imaging of both the stimulus and the magnetic response using simple uniform width channels. With a nontrivial cross channel geometry, the RF current and thus the FMR response are strongly nonuniform.

\section*{Acknowledgments}

This work was supported by AFOSR, under contract No. FA9550-14-1-0243. This work made use of the Cornell Center for Materials Research Shared Facilities which are supported through the NSF MRSEC program (DMR-1120296) as well as the Cornell NanoScale Facility, a member of the National Nanotechnology Infrastructure Network, supported by the NSF (Grant ECCS-0335765).

\bibliography{TRANE}

\end{document}